\documentclass[runningheads]{llncs}
\usepackage[T1]{fontenc}
\usepackage{graphicx,verbatim}
\usepackage{amsmath}
\usepackage{subcaption}
\usepackage{booktabs}
\usepackage{hyperref}       
\usepackage{xcolor}
\usepackage{sidecap}
\hypersetup{
    colorlinks,
    linkcolor={red!50!black},
    citecolor={blue!50!black},
    urlcolor={blue!50!black}
}
\usepackage{amsfonts}    

\begin{document}
\title{A Comprehensive Pipeline for\\Aortic Segmentation and Shape Analysis}
%
\author{Nairouz Shehata\inst{1,2} \and 
Amr Elsawy\inst{2} \and 
Mohamed Nagy\inst{2} \and
Muhammad ElMahdy\inst{3} \and
Mariam Ali\inst{4} \and
Soha Romeih\inst{5} \and 
Heba Aguib\inst{6} \and
Magdi Yacoub\inst{7,8} \and 
Ben Glocker\inst{1}} 
\authorrunning{N. Shehata et al.}
%
\institute{Department of Computing, Imperial College London, UK \and Biomedical \& Engineering Innovation Laboratory, Aswan Heart Centre, Egypt \and 
Department of Computer Science \& Engineering,\\The American University in Cairo, Egypt \and
Medical Physics \& Biomedical Engineering, University College London, UK \and Radiology Department, Aswan Heart Centre, Egypt \and
BMW Foundation Herbert Quandt, Germany \and Department of Surgery, Aswan Heart Centre, Egypt \and
National Heart \& Lung Institute, Imperial College London, UK\\
Corresponding author: \email{nairouz.shehata@myf-egypt.org}}


\maketitle              
\begin{abstract}
Aortic shape analysis plays a key role in cardiovascular diagnostics, treatment planning, and understanding disease progression. We present a robust, fully automated pipeline for aortic shape analysis from cardiac MRI, combining deep learning and statistical techniques across segmentation, 3D surface reconstruction, and mesh registration. We benchmark leading segmentation models—including nn-UNet, TotalSegmentator, and MedSAM2—highlighting the effectiveness of domain-specific training and transfer learning on a curated dataset. Following segmentation, we reconstruct high-quality 3D meshes and introduce a DL-based mesh registration method that directly optimises vertex displacements. This approach significantly outperforms classical rigid and non-rigid methods in geometric accuracy and anatomical consistency. Using the registered meshes, we perform statistical shape analysis on a cohort of 599 healthy subjects. Principal Component Analysis reveals dominant modes of aortic shape variation, capturing both global morphology and local structural differences under rigid and similarity transformations. Our findings demonstrate the advantages of integrating traditional geometry processing with learning-based models for anatomically precise and scalable aortic analysis. This work lays the groundwork for future studies into pathological shape deviations and supports the development of personalised diagnostics in cardiovascular medicine.


\end{abstract}
\section{Introduction}
Accurate aortic shape analysis is crucial in cardiovascular research for disease diagnosis, intervention planning, and monitoring disease progression. 
While 3D aorta segmentation has been well-studied in CT \cite{mavridis2024automatic,sedghi2019automated}, MRI remains challenging due to low resolution, weak contrast between the aorta and surrounding tissues, and intensity inhomogeneities \cite{ayed2014tric,heinrich2017mri}. This difficulty was also reflected in the Multi-Modality Whole Heart Segmentation Challenge (MM-WHS), where dice scores for aorta segmentation were consistently lower in MR compared to CT images \cite{zhuang2019evaluation}. Deep Learning (DL) and atlas-based approaches \cite{hepp2020fully} have partially mitigated these challenges but still require careful post-processing to ensure anatomical accuracy and topological consistency. Often, 3D surface reconstruction follows segmentation. Classical methods, such as Poisson Surface Reconstruction \cite{kazhdan2006poisson}, the Ball-Pivoting Algorithm \cite{bernardini1999ball}, and others \cite{boissonnat1993three,fortune2017voronoi}, assume smoothness and even sampling. If these assumptions are not met, they can produce holes or incorrect connections \cite{kulawiak2020improving}. As a result, surface reconstruction remains a central problem that current research is actively trying to solve, primarily through DL and hybrid approaches that are less reliant on rigid geometric assumptions \cite{liao2018deep,farshian2023deep,hanocka2020point2mesh}. 
Beyond segmentation and reconstruction, mesh registration is the final critical step for establishing dense anatomical correspondences across subjects—a prerequisite for statistical shape analysis. Registration methods can be broadly categorised into rigid and non-rigid approaches. Rigid methods \cite{paul1992besl,chen1992object,umeyama1991least} 
align shapes via rotation and translation, while non-rigid methods \cite{myronenko2010point,bookstein1991thin}
enable local deformations to better capture anatomical variability. These registration methods are sensitive to noise and initialisation \cite{amberg2007optimal}, prompting interest in DL-based approaches \cite{zhang2024comprehensive} that learn robust shape correspondences. 

We address these challenges by developing an integrated pipeline that combines statistical and DL methods for segmentation, surface reconstruction, and registration. Specifically, we evaluate state-of-the-art segmentation models including nn-UNet \cite{isensee2021nnu}, TotalSegmentator \cite{d2024totalsegmentator}, and MedSAM2 \cite{ma2024segment}—on a cardiac MRI dataset \cite{aguib2019genomics,aguib2020egyptian}. Following segmentation, we perform iso-surface extraction via Marching Cubes \cite{lorensen1987marching}, mesh post-processing for uniform point cloud densities, and apply advanced registration techniques to ensure accurate vertex correspondence. For the latter, we introduce a novel DL-based 3D aortic mesh registration method, which outperforms rigid (ICP \cite{paul1992besl,chen1992object}, RANSAC with FPFH \cite{zhou2016fast}) and non-rigid (CPD \cite{myronenko2010point} and Deformetrica \cite{bone2018deformetrica}) methods across all evaluation metrics. Finally, we perform statistical shape analysis using PCA to identify key modes of aortic shape variation in a cohort of 599 healthy subjects.

\section{Materials and Methods}
\textbf{Dataset.} We utilise a dataset of 744 cardiac MRI scans from subjects with standard aortic shapes, acquired using a Siemens Magnetom Aera 1.5 Tesla scanner at 'Anonymised Site' using the tfi3D-fs-free breathing navigator sequence. Images had 2.0mm pixel spacing and 1.6mm slice thickness. Ground-truth annotations were created for 200 scans by two expert annotators using Mimics Research version 19.0
, focusing on the thoracic aorta down to the diaphragm level. 
Preprocessing included clipping intensity values to the 0.5 and 99.5 percentiles to remove outliers and rescaling the intensity values to a standard range of 0-255 while preserving zero values. Images were aligned to the left posterior superior (LPS) coordinate system for MedSAM2 \cite{ma2024segment} and right anterior superior (RAS) coordinate system for nn-UNet \cite{isensee2021nnu} to ensure consistency.

\textbf{Segmentation.} We evaluated five models for segmenting the aorta from MRI: nn-UNet \cite{isensee2021nnu}, TotalSegmentator \cite{d2024totalsegmentator}, MedSAM2 \cite{ma2024segment}, and two fine-tuned versions of TotalSegmentator (\texttt{FT\_TS} and \texttt{FT\_TS\_NOFREEZING}). \texttt{FT\_TS} froze early encoder layers during fine-tuning, leveraging pre-trained feature extraction, while \texttt{FT\_TS\_NOFREEZING} allowed all layers to adapt to the dataset. We trained the models using 5-fold cross-validation on the 200 annotated scans. The best-performing model is used for inference and quality-controlled manual inspection.

\begin{figure}[h]
    \centering
	\includegraphics[width=0.9\textwidth]{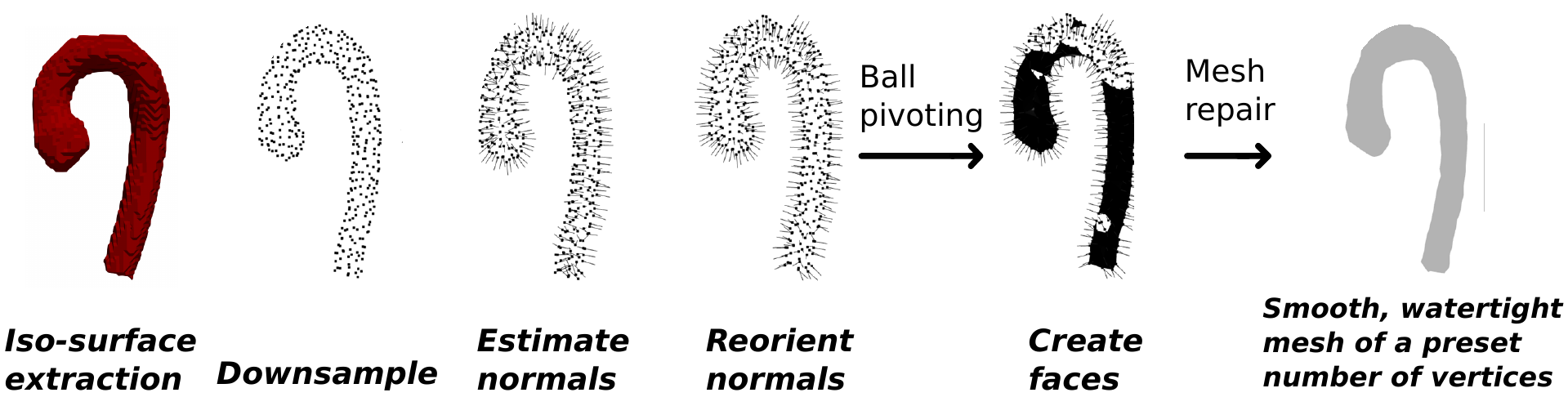}
	\caption{Overview for 3D surface reconstruction steps following segmentation. \label{fig:meshpro}}
\end{figure}

\textbf{3D Surface Reconstruction.} The process shown in Fig. \ref{fig:meshpro} starts with iso-surface extraction using the Marching Cubes algorithm \cite{lorensen1987marching}, producing initial meshes of the aorta. To standardise the number of vertices, we apply Poisson disk sampling \cite{kazhdan2006poisson} with a target count of 1000, ensuring approximately equidistant vertices (with minor variations due to the method’s stochasticity). After downsampling, normals are estimated by fitting local planes \cite{pauly2003point} and reoriented via tangent plane propagation \cite{zhou2018open3d} to prevent issues like inward-pointing normals or holes. The Ball Pivoting algorithm \cite{bernardini1999ball} then reconstructs mesh faces using multiple radii to handle varying point densities. We apply pymeshfix\footnote{\url{https://pymeshfix.pyvista.org/}} to create watertight manifold surfaces, addressing holes, non-manifold edges, and degenerate faces, while discarding disconnected components and inverting disoriented faces. Smoothing is also performed to enhance mesh quality and remove artefacts. Finally, to ensure a consistent vertex count across all subjects, additional vertices are added at the midpoints of the longest edges, bringing each mesh to precisely 1000 vertices. We selected 1,000 as a practical trade-off: it is high enough to preserve key anatomical features while keeping registration and PCA computation tractable, and ensures uniform sampling independent of the template’s resolution.

\begin{figure}[h!]
\centering
\begin{subfigure}[b]{0.49\textwidth}
\centering
\includegraphics[width=\textwidth, trim=2cm 0cm 0cm 0cm, clip]{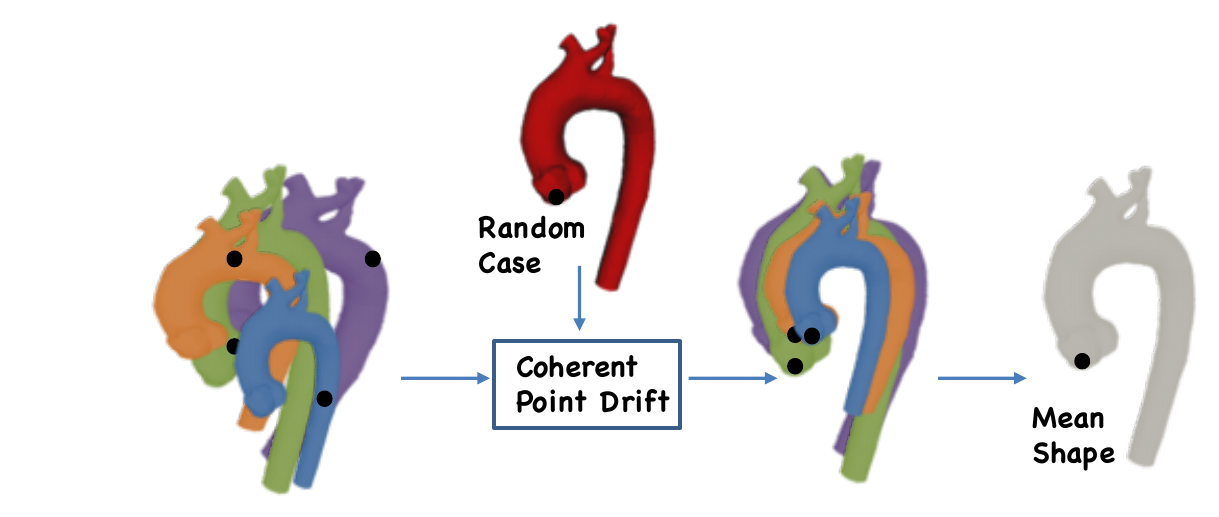}
\end{subfigure}
\hfill
\begin{subfigure}[b]{0.49\textwidth}
\centering
\includegraphics[width=\textwidth,trim=1cm 0cm 0cm 0cm, clip]{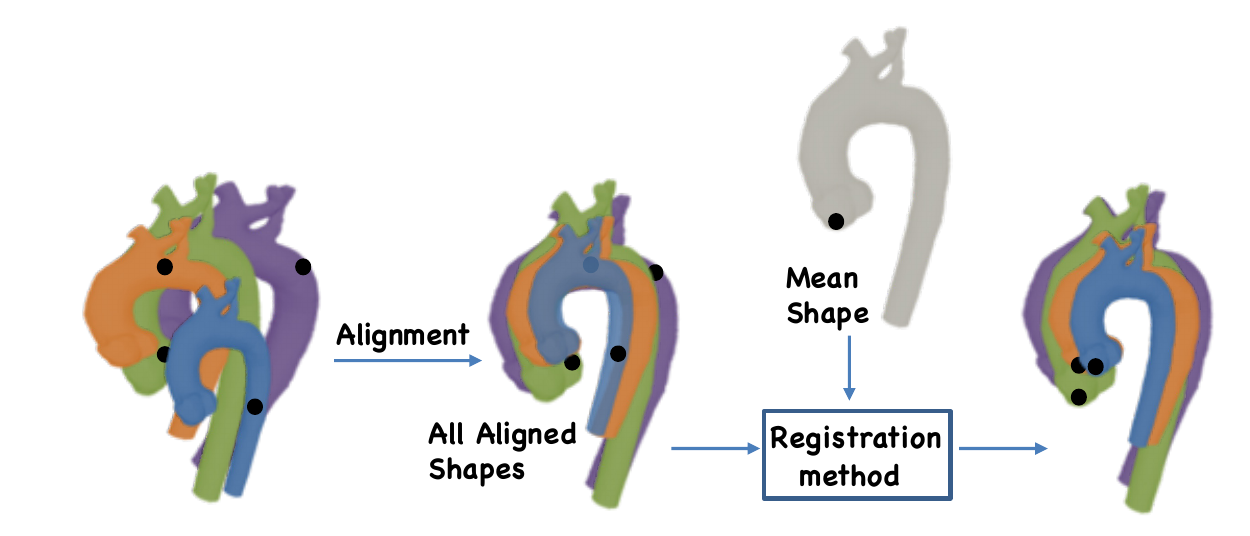}
\end{subfigure}
\caption{Left: Mean shape generation from K shapes via CPD affine and averaging. Right: 
Correspondence established by registering all meshes to template.}
\label{fig:meshreg}
\end{figure}

\textbf{Template Selection and Alignment.} 
A population-level template mesh was constructed by averaging \(K = 100\) randomly selected meshes, each registered to a randomly chosen initial mesh using affine CPD. The resulting mean shape serves as a neutral reference geometry, reducing anatomical bias, providing a canonical representation for downstream analysis and enabling consistent correspondence across the dataset (Fig.\ref{fig:meshreg}). We computed translation vectors to align the centroids of all meshes to the template mesh, ensuring spatial normalisation. Subsequent registrations deform this template to match individual target meshes and are evaluated using root mean square error (RMSE), Haussdorff, IoU, Dice, and the average distance between the target and deformed source (template) mesh.

\textbf{Baseline Mesh Registration Methods.}  
We deform the template (\textit{source}) to match each individual mesh (\textit{target}) in the dataset \(\mathcal{M}_{\text{src}}\) and \(\mathcal{M}_{\text{tgt}}\). We evaluated several registration methods, ICP (Point-to-Point \cite{paul1992besl} and Point-to-Plane \cite{chen1992object}), RANSAC with Fast Point Feature Histogram (FPFH) features \cite{zhou2016fast}, also followed by refinement with ICP (as RANSAC+ICP) and CPD affine and deformable registration \cite{myronenko2010point}. Deformetrica was used with a Large Deformation Diffeomorphic Metric Mapping (LDDMM) framework \cite{bone2018deformetrica}, employing Gaussian kernels with \(\lambda_{V}=10\,\text{mm}\) for deformation stiffness,\(\lambda_{W}=8\,\text{mm}\) for geometric detail, and a 1mm noise term to preserve anatomy. Optimisation was performed for 100 iterations using L-BFGS with automatic differentiation, starting from a 2mm (edges) isotropically remeshed template \cite{Sultan2025ImageBased}.
For efficient nearest-neighbour search and to maintain consistent topology, a KD-tree was employed to reorder target points in alignment with the transformed source. All methods are implemented using Open3D \cite{zhou2018open3d} and PyCPD \cite{gatti2022pycpd} for CPD. 


\textbf{Proposed Learning-based Mesh Registration.} 
This method is implemented using PyTorch3D \cite{ravi2020pytorch3d}\footnote{\url{https://pytorch3d.org/}}. We formulate registration as a vertex-wise deformation problem. Given a source mesh \( \mathcal{M}_{\text{src}} = (\mathcal{V}_{\text{src}}, \mathcal{F}_{\text{src}}) \) and a target mesh \( \mathcal{M}_{\text{tgt}} \), we learn a displacement field \( \Delta \mathcal{V} \in \mathbb{R}^{N \times 3} \) such that the deformed mesh \( \mathcal{M}_{\text{def}} = (\mathcal{V}_{\text{src}} + \Delta \mathcal{V}, \mathcal{F}_{\text{src}}) \) aligns closely with the target while preserving geometric and topological structure. 
In contrast to common DL-based registration approaches that employ convolutional or graph-based architectures to predict displacements \cite{litany2017deep,wang20193dn}, we directly optimise vertex positions via gradient descent. The optimisation minimises a composite loss function \(\mathcal{L}_{\text{total}}\) that captures global alignment, local smoothness, and anatomical plausibility:
\[
\mathcal{L}_{\text{total}} = \sum_{i=1}^{7} w_i \mathcal{L}_i + \lambda_{\text{reg}} \sum_{i=1}^{7} w_i \log(w_i + \epsilon)
\]

where \(\mathcal{L}_i\) denotes the individual loss terms, \( \lambda_{\text{reg}} \) a regularisation factor that penalises degenerate weight distributions and \(w_i\) are learnable, normalised weights constrained to be positive (sum to one) using a softmax over their log-space representation. This log-space parameterisation ensures stability and meaningful scaling. Additionally, weight regularisation prevents the collapse of the loss function to a single dominant term:
\[
w_i = \frac{\exp(\log w_i)}{\sum_{j=1}^{7} \exp(\log w_j)}
\]

The loss terms are as follows:
\begin{itemize}
    \item \textbf{Chamfer Distance}: Encourages pointwise correspondence between the deformed and target vertices.
    \item \textbf{Edge Length Loss}: Penalises deviations from original edge lengths to preserve mesh topology.
    \item \textbf{Normal Consistency}: Promotes alignment of surface normals to ensure smooth geometric transitions. 
    \item \textbf{Laplacian Smoothing}: Minimises Laplacian energy to reduce high-frequency surface noise in the deformed surface.
    \item \textbf{Curvature Matching}: Aligns intrinsic surface geometry by minimising the Chamfer distance between mean curvature values computed via the Laplacian operator.

    \item \textbf{IoU and Dice Coefficients}: Enforce volumetric consistency and spatial overlap between the deformed and target meshes.
\end{itemize}


We jointly optimise the vertex displacements \( \Delta \mathcal{V} \) and the log space loss weights \( \log w_i \) using the Adam optimiser. We use a two-tier learning rate: 1.0 for vertex displacements and 0.01 for log-weights. Each iteration involves:
\begin{enumerate}
    \item Computing the deformed mesh \( \mathcal{M}_{\text{def}} \). 
    \item Evaluating each loss component \( \mathcal{L}_i \) and computing their weighted sum.
    \item Updating parameters via backpropagation.
    \item Early stopping if the total loss fails to improve for 200 consecutive iterations.
\end{enumerate}


\textbf{Shape Analysis using Principal Component Analysis.} Out of 744 samples, 625 were self-admitted volunteers and 119 presented with complaints. The latter were excluded to focus strictly on healthy subjects. After further curation, 599 of the 625 volunteers were labelled healthy. PCA was performed on the registered meshes of these 599 healthy subjects, comparing Umeyama rigid and similarity transformations \cite{umeyama1991least} to distinguish global size effects from localised shape variation.

\section{Results and Discussion}
\begin{figure}[h!]
\centering
\begin{subfigure}[b]{0.48\textwidth}
\centering
\includegraphics[width=\linewidth]{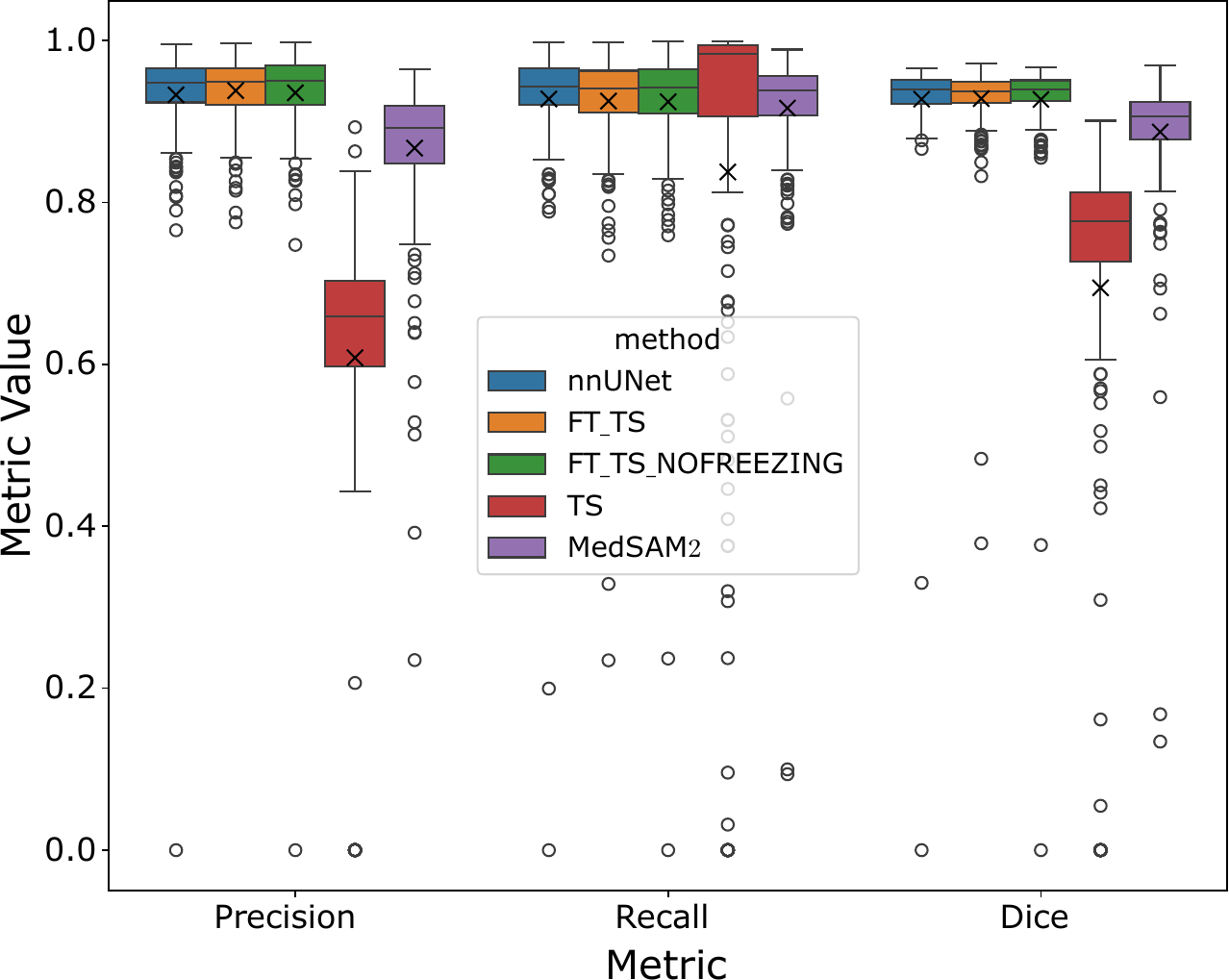}
\caption{Segmentation performance metrics}
\label{fig:seg_dice_comp}
\end{subfigure}
\hfill
\begin{subfigure}[b]{0.48\textwidth}
\centering
\includegraphics[width=\linewidth, trim = 21cm 1.4cm 17cm 1.4cm, clip]{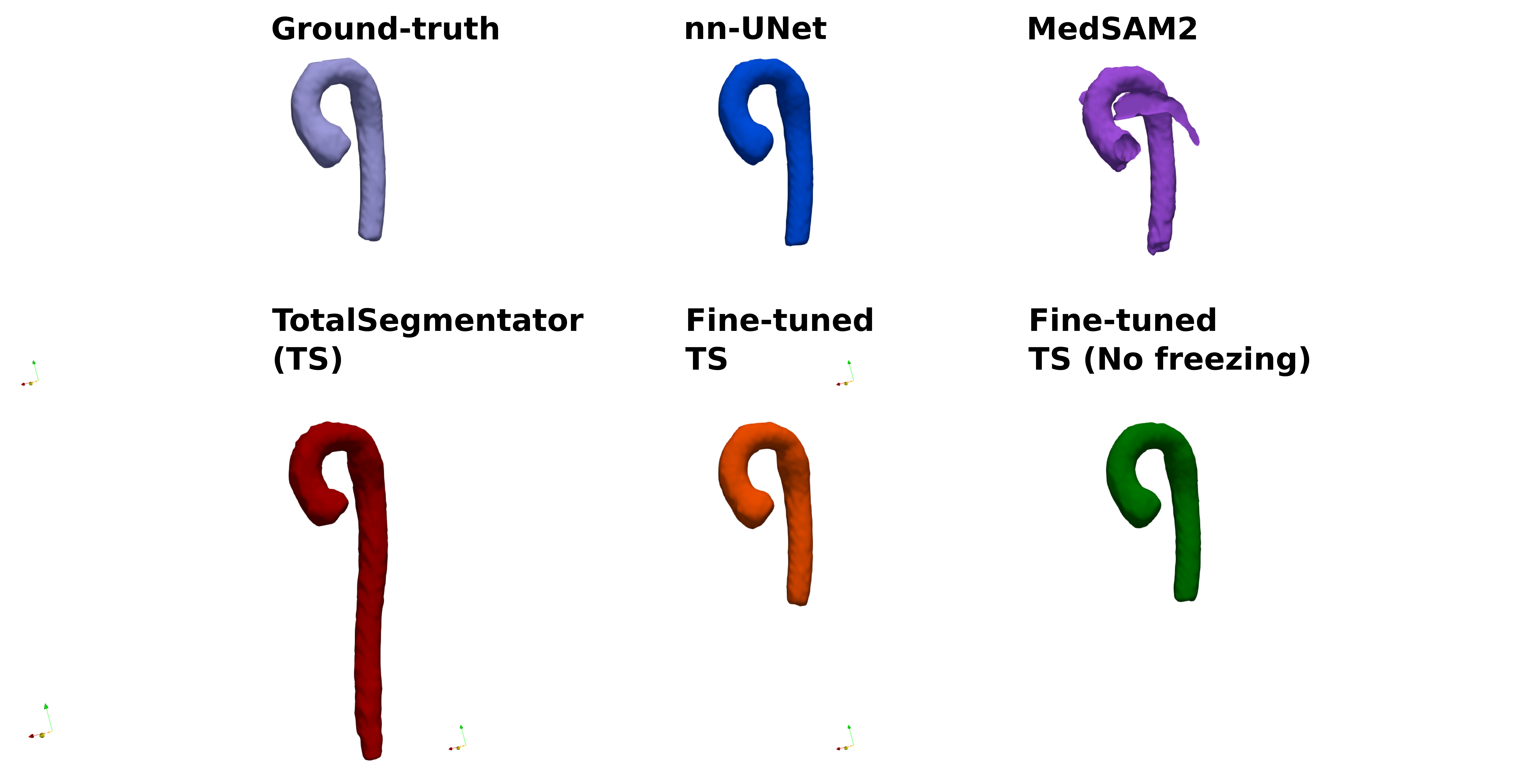}
\caption{Segmented random sample}
\label{fig:seg_example}
\end{subfigure}
\caption{Segmentation results.}
\label{fig:segmentation_results}
\end{figure}

\textbf{Segmentation Performance.} Our results in Fig. \ref{fig:segmentation_results} highlight the trade-offs between pre-trained, fine-tuned, and fully trained segmentation models. Non-fine-tuned TotalSegmentator oversegmented due to discrepancies between its pre-training data and our annotations. Fine-tuned MedSAM2 performed comparably but showed tendencies toward oversegmentation, suggesting potential improvements through post-processing. Comparing \texttt{FT\_TS} and \texttt{FT\_TS\_NOFREEZING} showed freezing layers had minimal impact on performance, suggesting unfrozen layers adapt well to the task-specific dataset. However, nn-UNet, trained from scratch, achieved the best overall performance, emphasising the value of task-specific optimisation despite higher computational costs. nnU-Net was selected for inference, but failed with 18 test images due to acquisition issues such as signal loss in the ascending aorta or aortic arch, incomplete aorta capture, and image noise. These difficulties underscore the importance of addressing data quality in real-world applications.

\begin{figure}[h!]
    \centering
	\includegraphics[width=\textwidth, trim = 5cm 0cm 5cm 0cm, clip]{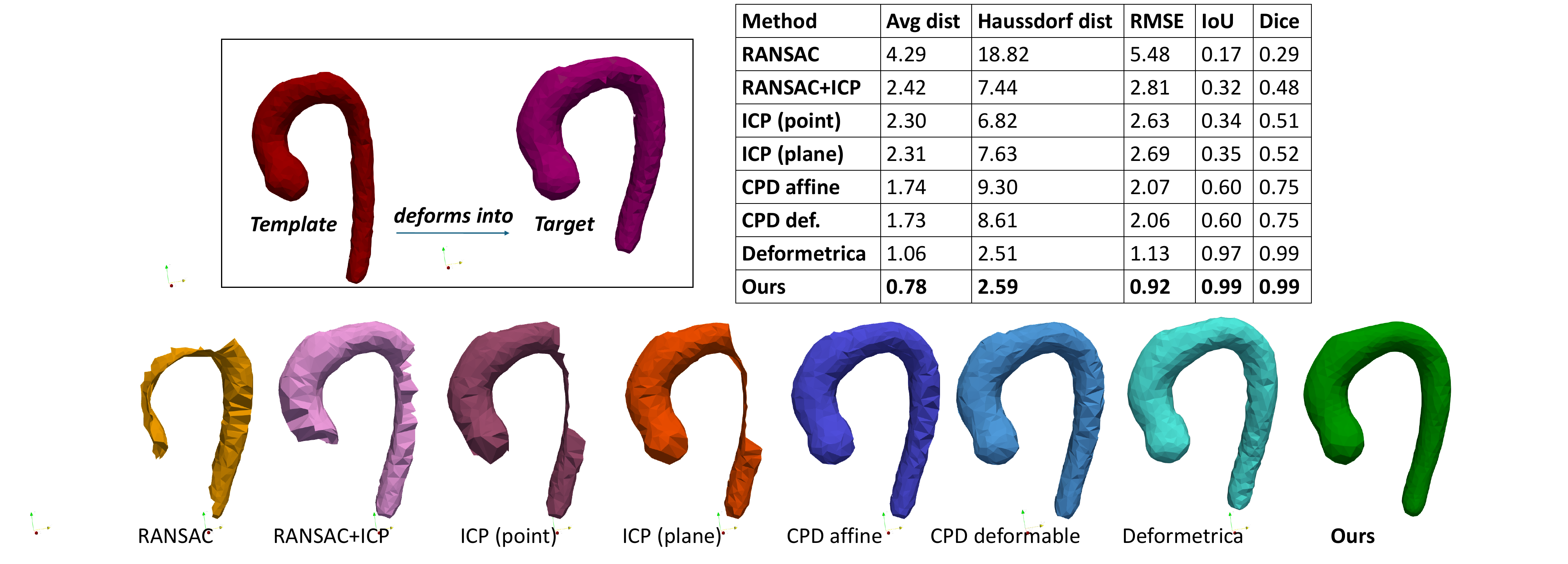}
	\caption{Random mesh used to compare how different methods deform template. \label{fig:reg_sample}}
\end{figure}

\textbf{Registration Performance.} 
The outlier rejection mechanisms of rigid registration methods, while designed for robustness, often failed in our setting due to their assumption of purely global transformations and inability to model localised shape differences. This led to misalignment in anatomically variable regions, causing partial overlap, artificial holes, and incomplete template coverage. In contrast, non-rigid approaches performed better by allowing pointwise deformation. CPD affine (global scaling and shearing) aided coarse alignment, but lacked local deformation modelling and is sensitive to initialisation, limiting fine-scale accuracy. CPD deformable improves flexibility through motion coherence but offers limited anatomical regularisation and struggles with ambiguous correspondences. Deformetrica produced smooth, invertible deformations with strong anatomical plausibility, explaining its high performance despite greater computational cost and parameter sensitivity. Our method outperformed all baselines across evaluation metrics (Table \ref{tab:registration}) by directly optimising vertex displacements via a composite loss that captures both global alignment and fine-grained geometry. This preserved mesh topology, enforced smoothness, and adapted to anatomical variability, yielding the most accurate and consistent registrations across the cohort.
Visual comparison in Fig. \ref{fig:reg_sample} reinforces these findings.
\begin{table}[h!]
\centering
\caption{Evaluation of registration methods (mean ± standard deviation)}
\label{tab:registration}
\begin{tabular}{lccccc}
\toprule
Method & Avg. Dist $\downarrow$ & Hausdorff $\downarrow$ & RMSE $\downarrow$ & IoU $\uparrow$ & Dice $\uparrow$ \\
\midrule
RANSAC & 3.94 $\pm$ 2.47 & 13.64 $\pm$ 9.11 & 4.95 $\pm$ 3.39 & 0.26 $\pm$ 0.14 & 0.40 $\pm$ 0.17 \\
RANSAC+ICP & 2.51 $\pm$ 1.91 & 10.02 $\pm$ 7.09 & 3.16 $\pm$ 2.49 & 0.46 $\pm$ 0.17 & 0.61 $\pm$ 0.16 \\
ICP (Point) & 2.31 $\pm$ 0.89 & 8.93 $\pm$ 4.92 & 2.81 $\pm$ 1.30 & 0.44 $\pm$ 0.17 & 0.60 $\pm$ 0.16 \\
ICP (Plane) & 2.28 $\pm$ 0.92 & 9.19 $\pm$ 5.42 & 2.82 $\pm$ 1.40 & 0.46 $\pm$ 0.16 & 0.61 $\pm$ 0.15 \\
CPD (Affine) & 1.45 $\pm$ 0.31 & 4.71 $\pm$ 2.08 & 1.60 $\pm$ 0.39 & 0.74 $\pm$ 0.17 & 0.84 $\pm$ 0.13 \\
CPD (Deformable) & 1.44 $\pm$ 0.30 & 4.77 $\pm$ 2.20 & 1.59 $\pm$ 0.38 & 0.75 $\pm$ 0.16 & 0.85 $\pm$ 0.12 \\
Deformetrica & 1.00 $\pm$ 0.19 & 2.63 $\pm$ 1.91 & 1.10 $\pm$ 0.41 & 0.97 $\pm$ 0.05 & 0.98 $\pm$ 0.03 \\
\textbf{Ours} & \textbf{0.73 $\pm$ 0.12} & \textbf{2.28 $\pm$ 0.55} & \textbf{0.85 $\pm$ 0.12} & \textbf{0.99 $\pm$ 0.03} & \textbf{0.99 $\pm$ 0.02} \\
\bottomrule
\end{tabular}
\end{table}

\begin{figure}[!h]
    \centering
    \begin{subfigure}[b]{0.98\textwidth}
        \includegraphics[width=\textwidth, trim=4cm 1.5cm 4cm 0cm, clip]{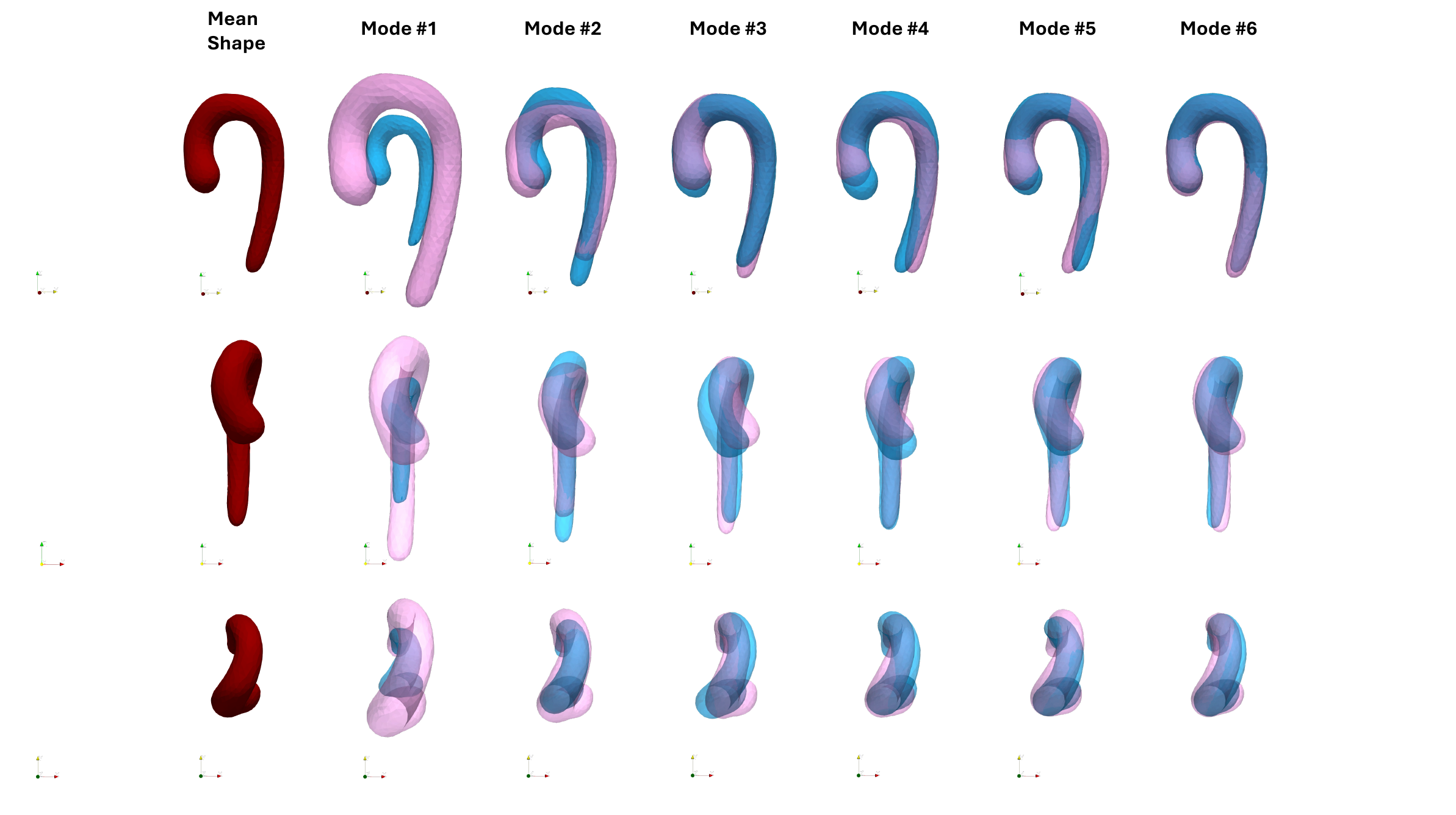}
        \caption{Umeyama rigid before applying PCA}
        \label{fig:pca_umrig}
    \end{subfigure}
    \hspace{0.02\textwidth} 
    \begin{subfigure}[b]{0.98\textwidth}
        \includegraphics[width=\textwidth, trim=4cm 1.5cm 4cm 0cm, clip]{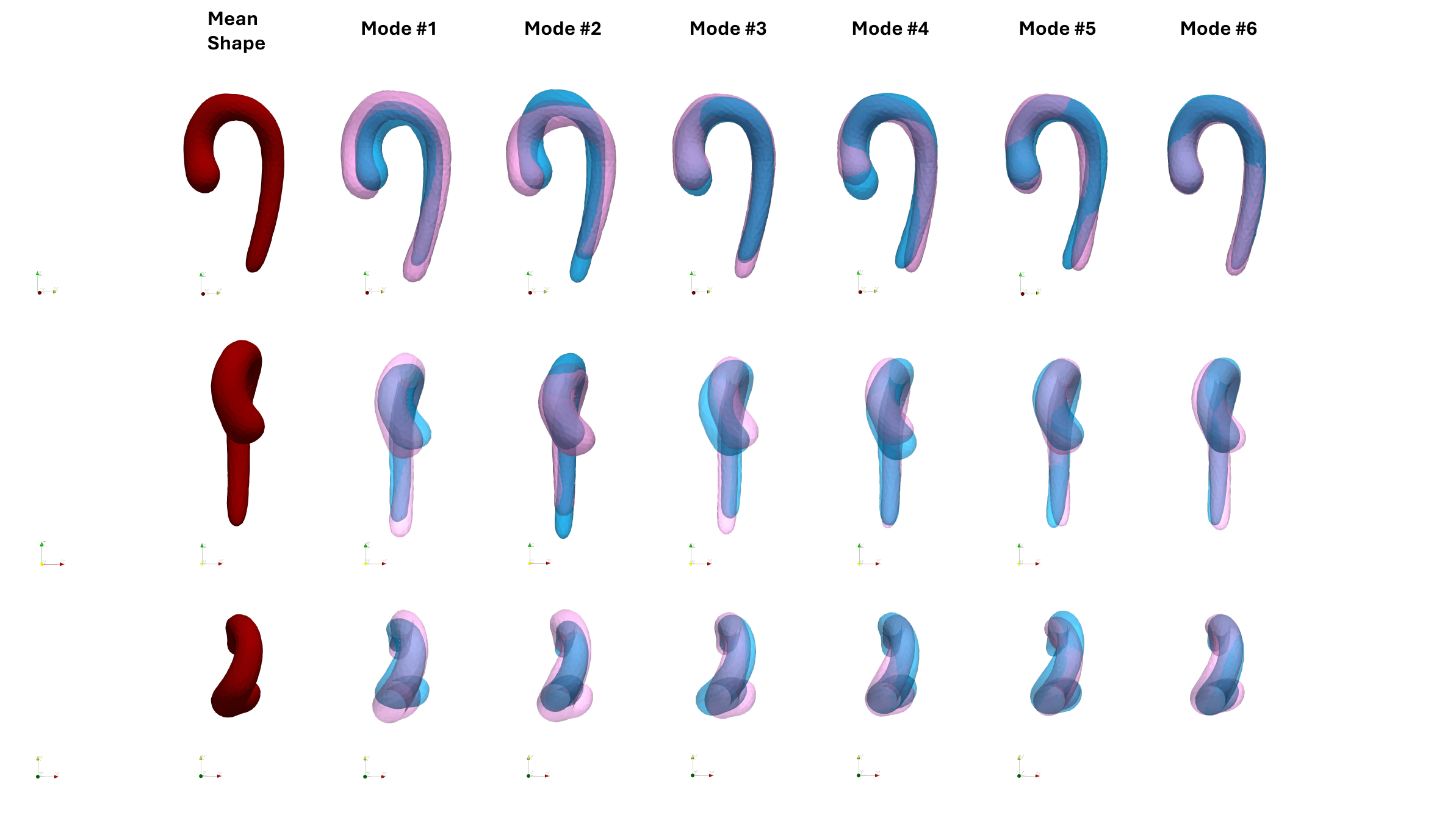}
        \caption{Umeyama similarity before applying PCA}
        \label{fig:pca_umsim}
    \end{subfigure}

    \caption{The panels from left to right show the average shape in bright red, followed by the top six modes of variation.The shapes after subtracting and adding two standard deviations of each mode to the average shape (in pink and blue, respectively). The three rows show the same shapes, from different viewing angles. (1) overall aortic size, (2) aortic arch height and width changes, (3) orientation of the ascending aorta, (4) relationship between arch curvature and ascending aorta length, (5) arch shape variation without length change, and (6) minor in-plane orientation adjustments. These findings underscore the importance of size-independent representations in shape analysis and provide insights into normal anatomical variability and potential pathological changes. \label{fig:pca_modes}}
\end{figure}

\textbf{Shape Analysis.} 
PCA results revealed distinct patterns of variance distribution under rigid and similarity transformations. The rigid transformation emphasised global size changes, with the first mode explaining 68.0\% of the variance, while the similarity transformation enabled a more detailed analysis of localised shape variations, with the first mode accounting for only 22.9\%. This redistribution leads to subsequent modes explaining relatively higher proportions of variance under similarity transformation, with consistent interpretations of shape characteristics across both approaches. The first six PCA modes capture key features of aortic shape variation described in Fig. \ref{fig:pca_modes}.



\section{Conclusion and Future Work}
We present a robust, automated pipeline for aortic segmentation, registration, and statistical shape analysis in healthy cardiac MRI using DL and statistical techniques. Our results demonstrate the effectiveness of fine-tuning and transfer learning for aortic segmentation, with nn-UNet achieving the highest Dice scores, while our DL-based registration method outperformed traditional approaches in both alignment and geometric fidelity, enabling precise statistical shape analysis. A key limitation is the restriction to healthy subjects, which constrains generalizability. However, our work establishes a highly accurate healthy reference model of aortic shape variation and identifies principal modes of morphological change, providing a baseline for future detection of pathological deviations. In addition, the resulting watertight 3D models produced by our pipeline are suitable for downstream applications such as computational fluid dynamics (CFD) simulations, enabling future integration with hemodynamic analyses. Future work should expand this approach to larger, more diverse datasets (diseased) and investigate correlations between shape modes and clinical outcomes, aiming to validate these features as predictive markers for aortic disease. Overall, our pipeline offers a reproducible technical foundation and a valuable healthy reference cohort, advancing the field toward more precise and personalised aortic assessment.

\begin{credits}
\subsubsection{\ackname} This PhD was funded by a joint scholarship from the Al Alfi Foundation and the Magdi Yacoub Heart Foundation.

\subsubsection{\discintname}
B.G. is part-time employee of DeepHealth. Other authors have no competing interests to declare that are relevant to the content of this article.
\end{credits}

%
%
%
\clearpage
\bibliographystyle{splncs04}
\bibliography{refs}

\end{document}